\documentclass[sigconf]{acmart}
\AtBeginDocument{%
  \providecommand\BibTeX{{%
    \normalfont B\kern-0.5em{\scshape i\kern-0.25em b}\kern-0.8em\TeX}}}

\usepackage{float}

\copyrightyear{2021}
\acmYear{2021}
\setcopyright{acmcopyright}
\acmConference[SIGIR '21]{Proceedings of the 44th International ACM SIGIR Conference on Research and Development in Information Retrieval}{July 11--15, 2021}{Virtual Event, Canada}
\acmBooktitle{Proceedings of the 44th International ACM SIGIR Conference on Research and Development in Information Retrieval (SIGIR '21), July 11--15, 2021, Virtual Event, Canada}

\copyrightyear{2021}
\acmYear{2021}
\setcopyright{acmlicensed}
\acmConference[SIGIR '21] {Proceedings of the 44th International ACM SIGIR Conference on Research and Development in Information Retrieval}{July 11--15, 2021}{Virtual Event, Canada.}
\acmBooktitle{Proceedings of the 44th International ACM SIGIR Conference on Research and Development in Information Retrieval (SIGIR '21), July 11--15, 2021, Virtual Event, Canada}
\acmPrice{15.00}
\acmISBN{978-1-4503-8037-9/21/07}
\acmDOI{10.1145/3404835.3463101}


\begin{document}
\fancyhead{} 
\title[Podcast Metadata and Content]{Podcast Metadata and Content: Episode Relevance and Attractiveness in Ad Hoc Search
}


\author{Ben Carterette$^{1}$, Rosie Jones$^{1}$, Gareth F. Jones$^{2}$, Maria Eskevich$^{3}$, Sravana Reddy$^{1}$, Ann Clifton$^{1}$, Yongze Yu$^{1}$, Jussi Karlgren$^{1}$, Ian Soboroff$^{4}$}
\affiliation{\institution{$^{1}$Spotify, $^{2}$Dublin City University, $^{3}$CLARIN ERIC, $^{4}$NIST} \country{$^{1}$United States and Sweden, $^{2}$Ireland, $^{3}$Netherlands, $^{4}$United States}}

\renewcommand{\shortauthors}{Carterette, et al.}

\begin{abstract}
Rapidly growing online podcast archives contain diverse content on a wide range of topics. These archives form an important resource for 
entertainment
and professional use, but their value can only be realized if users can rapidly and reliably locate content of interest. Search for relevant content can be based on metadata provided by content creators, but also on transcripts of the spoken content itself.
Excavating relevant content from deep within these audio streams for diverse types of information needs requires varying the approach to systems prototyping.
We describe a set of diverse podcast information needs and different approaches to assessing retrieved content for relevance. We use these information needs in an investigation of the utility and effectiveness of these information sources.
Based on our analysis, we recommend approaches for indexing and retrieving podcast content for ad hoc search.
\end{abstract}


\maketitle

\section{Introduction}

Podcasts are a rapidly expanding as a popular 
medium for delivery of spoken audio content.
As of 2021, more than 38 million podcast episodes are available online \cite{musicmph}.
Given the amount of podcast material available, 
we believe it is 
increasingly important that it be 
fully searchable if it is to be fully exploited by users. 

At present, most podcast search is done via {\em catalog match},
using show titles, episode titles, and sometimes metadata provided by podcast creators. This metadata is of highly varied quality and hence usefulness in supporting search operations.
Recommendation from friends and family \cite{Edison2019Consumer} remains in the top-three ways people find podcasts, while non-podcast-listeners in the same study say that they do not know how to find a podcast, or that they do not really know where to start.


In this paper we report experiments using the test collection from the TREC 2020 Podcasts Track which
show that the automatically transcribed content of podcast episodes is more reliably useful for search than metadata provided by podcast creators. 



\section{Podcast Format}


Podcasts are distributed as audio streams or files,
through RSS feeds
containing multiple metadata fields \cite{podcastersguiderss}.
A podcast {\em show\/} has a title, description, language, consumption order (episodic or sequential), and a list of categories (e.g., News, Sports, Comedy) selected by the creator from a predefined taxonomy.
A show typically has multiple {\em episodes}, which are the distinct audio  files.
Each episode has its own title, description, and other information.
All this metadata may be noisy or inadequate~\cite{sharpe2020review}.


Podcasting is 
typically a spoken-word medium.
However, the relative ease and low cost of recording and publishing means there is great  variability in the specifics.
Podcast episodes have a wide range of lengths, from just a few minutes to hours, although they tend to be 
between half an hour and an hour long.
Podcast content can vary in form, e.g. being scripted or informal dialogues. These can pose problems for search systems built for text archives, and also for the transcription of the content.


\section{Speech retrieval}
At first glance, podcasts are spoken documents, which have been well-studied 
in the TREC Spoken Document Retrieval Track which ran from 1997-2000 \citep{GarafoloEtAl2000}. However, this work was  based on news corpora, which are relatively homogenous in genre, style, and speaking professionalism, while podcasts come in many disparate forms, increasing the difficulty of the task.
Retrieval from an archive of oral history has also been well-studied  \citep{PecinaEtAl2008}, but lacks the multi-speaker, multi-genre aspect of podcasts. 
The NTCIR Spoken Query and Document tasks \citep{AkibaEtAl2016} used a document collection of spontaneous speech, but the 600 hours of speech is tiny compared to the 50,000 hours of speech in the Spotify podcasts corpus \cite{clifton2020hundredthousand}.
The identification of ``jump-in'' points in multimedia content based on the spoken soundtrack at  Mediaeval from 2011-2015 \citep{LarsonEtAl2011, EskevichEtAl2012, EskevichEtAl2015} motivates an approach to podcast search in which we identify the best place to start listening.
We refer to \citet{jones2019} for a more complete overview of research in spoken content retrieval from its beginnings in the early 1990s to today. 

Because of the availability of metadata fields,
podcasts can be represented as semi-structured documents, and 
models like BM25F~\cite{Robertson:2004}, Field Relevance Models~\cite{Kim:2012}, and NRMF~\cite{Zamani:2018:NRMF}, can be adopted for podcast search tasks. Evaluation campaigns such as the INEX XML retrieval initiative \cite{Fuhr:2006,Lalmas:2007} have studied such models.
As argued by \citet{Besser:2008:UserGoalsPodcastSearch}, the goals of podcast search may be similar to those for blog search, if podcasts are viewed as audio blogs. 

In addition, the publication format of podcasts, as series of episodes typically consumed in sequence, and the prominence of hosts and certain popular guests, act as a filter on top of the topical search. \citet{Manos:2008:PodCred} argued that the quality and credibility of podcasts, which are sometimes considered during the relevance assessment process, can be characterized using four types of indicators pertaining to the podcast content, the podcast creator, the podcast context, or the technical execution of the podcast. These facts distinguish podcast search from most well-established search tasks including adhoc, web, personal, and enterprise search.




\section{TREC 2020 Podcasts Track}

The Podcasts Track ran for the first time at TREC 2020 
~\cite{trec2020podcastnotebook}. It featured two tasks: a search task 
and a summarization task, the former is the focus of this paper.

Podcast search 
focused on a segment retrieval task 
defined as the problem of finding relevant segments of podcast episodes given a query representing a specific information need.
The corpus for the task comprised 100,000 podcast episodes released in 2019, including metadata, audio, and full transcripts produced by automatic speech recognition (ASR)~\cite{clifton2020hundredthousand}.
Participants were asked to retrieve unique episode identifiers (episode URIs) along with a time offset to the start of a two-minute segment starting on the minute within that episode.
The segment corpus comprises 3.4M segments with an average word count of $340 \pm 70$ per segment.



Seven participating groups submitted a total of 24 runs to the search task. 
Each run ranked up to 1,000 segments for each of the 50 queries.
Runs used retrieval techniques such as relevance feedback, query expansion, word2vec, BERT reranking, and fusion.

\section{Podcast search}

We now take a step back to ask: why are we interested in this kind of podcast search?
What are the user needs that can be addressed by retrieving segments of episodes?
What other information could be retrieved that might improve the user experience?

\subsection{Information needs for podcast search}

User needs pertaining to podcasts include  education, entertainment, and information.
The format and variability of the podcast medium -- series of episodes, the importance of the specific host or guest in the episode, range of presentation styles from monologues to interviews and banter, production quality   -- 
are additional aspects that may determine whether a user is interested in listening or not, entirely separate from topical relevance. These factors make podcast search different from traditional search tasks.


\subsection{Topic development for TREC}
\label{sec:topicdev}


Since there is currently no large-scale content-based podcast search engine, 
there is no simple source of sample topics.
Thus we 
developed topics by introspection, considering what we might use a podcast search engine for, what we could imagine others using it for, what types of information needs make it more attractive than web search, etc.
Topic development used 
 several sources:  lists of events in 2019, 
topic creators interests, and browsing metadata for potentially interesting content.  
To determine whether the topic would be interesting, we roughly compared metadata matches, web search results, and the results from our in-house transcript search engine.  
Topics were finally selected based on whether they retrieved interesting content from a simple search index of the podcast transcripts.
  
  We also experimented with ``known item'' and ``refinding'' information needs.
In total, eight development and 50 test topics were developed, with 13 topics within the test set being labeled as refinding or known-item.

\begin{figure}
    \centering
    \footnotesize
    \begin{verbatim}
<topic>
<num>34</num>
<query>halloween stories and chat</query>
<type>topical</type>
<description>I love Halloween and I want to hear stories and conversations 
             about things people have done to celebrate it.  I am not looking 
             for information about the history of Halloween or generalities 
             about how it is celebrated, I want specific stories from 
             individuals.
</description>
</topic>
\end{verbatim}
    \caption{Example search topics}
    \label{queryexample}
\end{figure}

%
%
%
%

\subsection{Assessing relevance}
\subsubsection{TREC}

Participant submissions were pooled to depth 20 (minimum 128 segments per topic, maximum 306) and reviewed by NIST assessors.  The assessors primarily reviewed the ASR text of the segment, but could listen to audio if the need arose.  The assessor’s view showed each retrieved segment within the context of the transcript of the full episode, so that they could explore the context of the segment to better understand it.  All segments from the same episode were judged in sequence together. 

The relevance judgments were on a four-point scale of bad (0), fair (1), good (2), or excellent (3). For known-item and refinding topics, an additional level of “perfect” (4) was added and meant that the segment was precisely what the user was looking for.  Excellent segments were completely on-topic, provided highly relevant information, and represented an ideal entry point into the episode.  

\subsubsection{Additional Assessments}
\begin{table}[tbp!]
    \centering
            \resizebox{\columnwidth}{!}{%
    \begin{tabular}{cclr}
source & no.\ topics & type & no.\ assessments \\
\hline
TREC & 16 & title & 1,381 \\
     &    & title+description & 1,381 \\
     &    & transcript-segment & 1,863 \\
     &    & transcript-full & 114 \\
     \hline
metadata & 50 & title & 1,049 \\
+transcript &  & title+description & 1,049 \\
indexes &  & transcript-segment & 514 \\
        &  & transcript-full & 254 \\
        \hline
    \end{tabular}}
    \caption{New assessments collected for this work. Assessments were collected from documents retrieved for a subsample of topics from TREC runs, and from all documents retrieved from indexes described in Section~\ref{sec:indexes}.}
    \label{tab:assessmentsummary}
\end{table}

The authors of this work independently assessed podcast metadata and full transcripts.
From the metadata, we extracted episode titles and descriptions.
Episode titles and descriptions are frequently too short or unspecified to accurately reflect episode relevance, so instead we assess how {\em attractive} they might be to a user with the stated information need.

We produced the following additional assessments:

\begin{itemize}
    \item \textbf{Title attractiveness.} Would a user with the given information need find the episode a good candidate to stream based solely on the title?
    \item \textbf{Title and description attractiveness.}  Would the user  find the episode a good candidate to stream based on the title and description together? 
    \item \textbf{Full transcript relevance.} Is the episode relevant to the information need (based on the full transcript)? 
\end{itemize}

We also used TREC segment-level judgments to obtain transcript-level judgments by taking the maximum relevance of any segment of an episode as the relevance of the transcript.
We refer to this as the \textbf{transcript-segment} judgment.
We acknowledge that these may be unreliable; it is possible that there is a segment of the episode more relevant than any seen by a TREC assessor and therefore our transcript-level judgment is low.

We selected documents to judge as follows:
First, we built five new indexes using different combinations of metadata and transcripts; see Section~\ref{sec:indexes} for details.
We retrieved the top 10 results for all TREC topics from each of these five indexes, pooled them, and judged all titles and descriptions for attractiveness, plus a select subset of transcripts for relevance.  This provided about 1,000 episode title and description attractiveness judgments, plus about 250 transcripts that had not previously been assessed for TREC.

We also pooled the top 10 results from all 24 TREC submitted runs.  We selected a random sample of 16 topics for judging title and description attractiveness in this pool.
We judged another 1,400 episode titles and descriptions in this set, as well as another 100 transcripts.
Episodes were ordered by a function of rank position in runs, which may produce ordering effects~\cite{eisenberg88}. We did not investigate this.
Table~\ref{tab:assessmentsummary} summarizes the new assessments.

There is very little overlap between episodes from our new indexes and episodes from TREC submitted runs.
Across all 50 topics, there are only 79 episodes that occur in the top 10 of both our new runs and the 24 TREC submissions---only $7.7\%$ of all episodes retrieved by our five new runs.
By focusing on metadata we retrieve many more potentially relevant episodes.



\section{Indexing and Searching Podcasts}\label{sec:indexes}

An RSS feed with podcast episode titles, descriptions, and other metadata, in addition to the audio file, includes a lot of information that could be indexed for retrieval.
In order to understand the relative utility of various free-text fields, we constructed Lucene indexes of episode titles, episode descriptions, titles and descriptions concatenated, full ASR transcripts, and transcripts concatenated with titles and descriptions.
We retrieved rankings of episode URIs for all 50 TREC topics from each of these five indexes.

We also obtained the TREC submitted runs.
In order to make the segment rankings comparable to URI rankings from our indexes, we compressed all retrieved segments from one URI to a single result for that URI, ranked at the position of the top-ranked segment, and then removed all subsequent mentions of that URI.

Table~\ref{tab:my_label} demonstrates how the choice of assessment and index impact retrieval results.
Rows correspond to retrieval from the indexes described above.
Columns correspond to the four different judgment types summarized in Table~\ref{tab:assessmentsummary}. Conclusions from this table:

\begin{table}[tbp!]
    \centering
        \resizebox{0.99\columnwidth}{!}{%
    \begin{tabular}{l|cccc}
       & \multicolumn{4}{c}{assessment type} \\ 
 \hline       &       & title +     & transcript- & transcript-      \\ 
index  & title & description & segment & full \\ 
\hline 
title  & 0.43  & 0.33  & 0.19  & 0.21 \\ 
description  & 0.40  & 0.51  & 0.24  & 0.45 \\ 
title+description  & 0.49  & 0.56  & 0.27  & 0.46 \\ 
full transcript  & 0.37  & 0.42  & 0.41  & 0.52 \\ 
transcript+title+description  & 0.43  & 0.51  & 0.45  & 0.61 \\ 
 \hline
    \end{tabular}}
    \caption{NDCG@10 results for five different indexes evaluated with four assessment types.}
    \label{tab:my_label}
    \vspace{-1em}
\end{table}

  \begin{table}[tbp!]
    \centering
    \resizebox{0.95\columnwidth}{!}{%
    \begin{tabular}{l|cccc}
       & \multicolumn{4}{c}{assessment type} \\ 
 \hline       &       & title +     & transcript- & transcript-    \\ 
run  & title & description & segment & full \\ 
\hline 
oudalab1  & 0.03  & 0.05  & 0.02  & 0.00 \\ 
hltcoe1  & 0.17  & 0.23  & 0.18  & 0.26 \\ 
hltcoe5  & 0.16  & 0.18  & 0.19  & 0.22 \\ 
hltcoe3  & 0.15  & 0.19  & 0.24  & 0.20 \\ 
hltcoe2  & 0.25  & 0.28  & 0.31  & 0.29 \\ 
LRGREtvrs-r\_3  & 0.20  & 0.29  & 0.32  & 0.37 \\ 
BERT-DESC-TD  & 0.20  & 0.27  & 0.33  & 0.41 \\ 
BM25  & 0.20  & 0.27  & 0.33  & 0.41 \\ 
BERT-DESC-Q  & 0.20  & 0.27  & 0.33  & 0.41 \\ 
RERANK-QUERY  & 0.20  & 0.27  & 0.33  & 0.41 \\ 
RERANK-DESC  & 0.20  & 0.27  & 0.33  & 0.41 \\ 
BERT-DESC-S  & 0.20  & 0.27  & 0.33  & 0.41 \\ 
LRGREtvrs-r\_2  & 0.24  & 0.30  & 0.35  & 0.36 \\ 
QL  & 0.21  & 0.26  & 0.37  & 0.38 \\ 
LRGREtvrs-r\_1  & 0.25  & 0.32  & 0.37  & 0.38 \\ 
hltcoe4  & 0.29  & 0.29  & 0.43  & 0.32 \\ 
run\_dcu1  & 0.33  & 0.37  & 0.44  & 0.44 \\ 
run\_dcu3  & 0.33  & 0.36  & 0.46  & 0.43 \\ 
UTDThesis\_Run1  & 0.28  & 0.32  & 0.46  & 0.28 \\ 
run\_dcu5  & 0.35  & 0.36  & 0.47  & 0.45 \\ 
run\_dcu2  & 0.33  & 0.37  & 0.48  & 0.44 \\ 
UMD\_IR\_run2  & 0.32  & 0.34  & 0.48  & 0.37 \\ 
run\_dcu4  & 0.35  & 0.37  & 0.49  & 0.45 \\ 
UMD\_ID\_run4  & 0.32  & 0.37  & 0.54  & 0.40 \\ 
UMD\_IR\_run1  & 0.29  & 0.37  & 0.55  & 0.35 \\ 
UMD\_IR\_run3  & 0.31  & 0.39  & 0.58  & 0.41 \\ 
UMD\_IR\_run5  & 0.35  & 0.42  & 0.58  & 0.40 \\ 
\hline
    \end{tabular}}
    \caption{Evaluation of TREC submitted runs and baselines by NDCG@10 with four new types of assessments.} 
    \label{tab:retrieval-results}
    \vspace{-1em}
\end{table}

\begin{table*}[tb!]
    \centering
\resizebox{1.6\columnwidth}{!}{%
    \begin{tabular}{l|rrr|rr|r}
	      &    \multicolumn{3}{c|}{transcript} &   \multicolumn{2}{c|}{description}  &  \% transcript words \\
	      &	  avg.\ length  & vocab size       &  avg.\ ratio &  avg.\ length  & vocab size       &  in description \\
	          \hline

True Crime                &  2200 &  870   &  0.48	& 	  100  &      71  &   4.3       \\
Religion and Spirituality &  2000 &  700   &  0.44	& 	   79  &      55  &   3.5       \\
Government                &  2500 &  860   &  0.41	& 	   82  &      60  &   3.1       \\
Business and Technology   &  2900 &  910   &  0.38	&	  141  &      94  &   4.1       \\
News and Politics         &  2800 & 1200   &  0.46	& 	  100  &      72  &   2.8       \\
History                   &  2300 & 1000   &  0.49	& 	   81  &      58  &   2.9       \\
Fiction                   &  1900 &  770   &  0.52 	&	   78  &      53  &   2.6       \\
    \hline
    \end{tabular}}
    \caption{Terminological coverage of description with respect to transcript, content words}
    \label{tab:terms}
\end{table*}

\begin{enumerate}
    \item A ranking of episode titles from a title-only index is more attractive than a ranking of episode titles from a description-only index, while a ranking of descriptions from a description-only index is more attractive than a ranking of titles from a description-only index, suggesting that titles and descriptions are not always signaling the same topical relatedness.
    \item The most attractive results are achieved by retrieving against an index of episode titles and descriptions.
    \item A ranking based on retrieval of transcripts returns many more relevant episodes, 
    though titles and descriptions are much less attractive.
    \item Indexing transcripts and metadata together provides the strongest relevance by either transcript judgment type with a relatively small decrease in attractiveness.
    \item Result attractiveness remains surprisingly low in all cases---many results will appear not relevant.
\end{enumerate}

From this we further conclude that there is a great deal of relevant material to be excavated from the transcripts that cannot be accessed via the metadata.
But enabling users to find that information via full-document search means that result presentation will be negatively impacted.
In other words, metadata search is inadequate for finding relevant information, but full-text search results in unappealing rankings.
Thus some form of query-biased summarization \cite{spinaQueryBiased2017} or passage retrieval is necessary for presenting results to users in an appealing way.




  

Table~\ref{tab:retrieval-results} shows the TREC runs evaluated by the four new assessment types.
We observe the following:
\begin{itemize}
    \item There is stronger correlation between different assessment types among TREC runs than among those in Table~\ref{tab:my_label}.
    \item By the transcript-segment judgments, our indexes are not competitive with the best TREC submitted runs.
    \item By the new transcript-level relevance assessments, all but one of our runs are better than the best TREC run. 
    \item Our runs are substantially better than TREC runs on title and description attractiveness.
\end{itemize}
Based on these findings, it could be argued that indexing metadata and transcripts together, then searching full transcripts rather than segments, is likely to provide the best overall user experience.



\section{Assessment agreement}

\noindent\textbf{Transcript relevance.}
Transcript-segment and transcript-full assessments agree that an episode is relevant in 71\% of cases. 
43\% of episodes have agreement on the exact grade of relevance.
This is a high level of agreement compared to other retrieval tasks~\cite{voorhees98}.


\noindent\textbf{Attractiveness and relevance.}
When either the title is judged attractive {\em or} the episode is judged relevant, in about 50\% of cases both are true.
This is a decent level of agreement for a retrieval task, but it means that episode title attractiveness is often not a very good predictor of episode relevance.
The results are very similar for description attractiveness.

\noindent\textbf{Interannotator agreement.}
URIs that overlapped between the five metadata runs and the TREC runs were independently judged by two different assessors.
Among episode titles that at least one assessor marked attractive, assessors agreed 88\% of the time.  
Similarly with descriptions, in 82\% of cases both assessors found the description attractive.
Agreement on exact grade of attractiveness was high as well: 72\% for titles and 68\% for descriptions.



\section{Topical focus}
Table~\ref{tab:terms} shows the coverage of content words for select categories of podcasts. Some genres have episodes with relatively high topical focus, which is reflected in the topical coverage of the descriptions. Each transcript and  description is filtered to only contain nouns, verbs, and adjectives; the table reports counts of these tokens. 
The average ratio of vocabulary size to length, a measure of topical variation within a text, varies across the categories. 
While all these scores are normal, the topical variation of "Fiction" podcasts is considerably wider than that of "Business and Technology" podcasts.  
The last column of the table demonstrates a difference in terminological coverage. This can be understood as a measure of the topical representativity of the descriptions with respect to the episode. 
A higher score will mean that the description represents more of the topical content. 
The score variation indicates a potential for determining the topicality of the episode or show, and thus the utility of using search technology optimised for topical retrieval and content analysis (as opposed to usage-based similarity measures).

\section{Conclusion}
It is not difficult to find information needs for podcast search such that highly relevant content is buried in episodes, its presence not indicated by either episode title or description.
Episode titles and descriptions that appear attractive may lead to irrelevant content and frustrated users.
Podcast search engines should index both metadata and episode content, and episode segments or query-biased summaries may be necessary to help users understand why retrieved episodes are relevant to their need.

\textbf{Acknowledgments} 
Thanks to the reviewers for their constructive suggestions, and to TREC participants and organizers for helping make the track a success.
Gareth Jones is partially supported by Science Foundation Ireland as part of the ADAPT Centre (Grant 13/RC/2106) at Dublin City University.

\clearpage

\bibliographystyle{ACM-Reference-Format}
\bibliography{podcast-segment-retrieval}


\end{document}